%% file: Preprint.tex
\documentclass[preprint]{aastex631}

\shorttitle{VY CMa SW Knots} 
\shortauthors{Humphreys et al.}

\begin{document}

\title{The Infrared-Bright SW Knots in the Complex Ejecta of VY CMa}

\correspondingauthor{Roberta Humphreys}
\email{roberta@umn.edu}

\author{Roberta M. Humphreys}
\affiliation{Minnesota Institute for Astrophysics,  University of Minnesota, Minneapolis, MN 55455, USA}

\author{Terry J. Jones}
\affiliation{Minnesota Institute for Astrophysics,  University of Minnesota, Minneapolis, MN 55455, USA}

\author{Kris Davidson}
\affiliation{Minnesota Institute for Astrophysics,  University of Minnesota, Minneapolis, MN 55455, USA}

\author{A. M. S. Richards}
\affiliation{Department of Astronomy University of Manchester, UK}

\author{R. Ravi}
\affiliation{Department of Chemistry and Biochemistry University of Arizona, USA}

\author{A. P. Singh}
\affiliation{Department of Chemistry and Biochemistry University of Arizona, USA}

\author{L. M. Ziurys}
\affiliation{Department of Astronomy; Department of Chemistry, and Steward Observatory University of Arizona, USA}

\begin{abstract}
	The red hypergiant VY CMa is remarkable for its very visible record of high mass loss events observed over the range of wavelengths from the optical and infrared to the submillimeter region with ALMA. The SW Clump or SW knots are unique in the ejecta of VY CMa. Except for the central star, they are the brightest sources of dusty infrared emission in its complex ejecta.  In this paper we combine the proper motions from the HST images, and infrared fluxes from 2 to 12 microns with the $^{12}$CO  images from ALMA to determine their ages and mass estimates. The SW knots were ejected more than 200 years ago with an active period lasting about 30 years, and with a total mass in the Clump $>$ 2 $\times$ 10$^{-2}$ M$_{\odot}$.
	
\end{abstract}


\section{The SW Knots in VY CMa's Record of High Mass Outflows} \label{sec:intro}

The role of mass loss and episodic high mass loss events is observed in massive stars across the upper HR Diagram. Mass loss alters their evolution and may determine  their eventual fate as supernovae or as a direct collapse to a black hole depending on their initial mass and mass loss history.  These high mass loss events are clearly visible, sometimes quite dramatically, for example, in eta Carinae \citep{Morse}, the ejecta of some Luminous Blue Variables \citep{Weis}, the yellow and red hypergiants, and most recently in the "great dimming" of Betelgeuse \citep{Montarges,Dupree}. A history of high mass loss events is rarely so visible as in the ejecta of the red hypergiant VY CMa \citep{Smith,RMH2007}.

\vspace{2mm} 

Optical and near-infrared images of VY CMa reveal prominent arcs and clumps of knots that are spatially and kinematically distinct from its diffuse ejecta. Proper motions and line of sight velocities show that they were ejected at different times over several hundred years, in dfferent directions and from separate regions on the star \citep{RMH2007,RMH2019,RMH2021}. The most recent events correspond with extended periods of activity with deep minima in VY CMa's historic light curve \citep{RMH2021,RMH2024}.

\vspace{2mm}

Our recent high resolution ALMA images \citep{Singh} revealed a more extended ejecta with three prominent large arcs not seen in the HST optical images or in the infrared. Earlier ALMA data \citep{Richards,OGorman} had also revealed an optically obscured massive clump close to the star plus three nearby smaller clumps in higher resolution data \citep{Kaminski}. With our new ALMA $^{12}$CO  data we showed that they are separate outflows that were ejected  during the same 30 yr active period in the early 20th century as similar knots just west of the star in the HST images \citep{RMH2024}

\vspace{2mm}

In this paper we take a closer look at the infrared-bright SW Clump or SW knots combining the HST optical images and the longer wavelength data from 2 to 12 $\mu$m with the molecular images from our ALMA program. What we now call the SW Clump or SW knots was first recognized as a  bright infrared source \citep{Cruz}  which looked  like a singular dusty clump to the SW of the central star, see Figure 1a. HST WFPC2 images \citep{Smith,RMH2007} however  revealed  separate knots and a clumpy arc (Figure 1b) which in our projected line of sight resembles a possible bubble or expanding ejecta from a common event.  Near and mid-infrared images \citep{Shenoy13,Gordon} from 1.2$\mu$m to 12$\mu$m (Figure 2) showed, somewhat remarkably, that the emission was primarily due to scattering even at 5 $\mu$m. Thermal emission from dust dominates its energy distribution emission at the longer wavelengths, from 9 to 12 $\mu$m.  \citet{RMH2021} demonstrated that the apparent spatial distribution of the SW knots and diffuse arc could be fit by an ellipse although,  the  motions of the individual knots did not support an expanding  bubble. Nevertheless, the infrared flux associated with the SW knots is due to a total mass of  at least 0.02M$_{\odot}$, dust plus gas \citep{RMH2021,RHJ}.

\vspace{2mm} 
In the next section we describe the $^{12}$CO images used in this discussion. In \S {3} we summarize  the motions and ages of the SW knots and compare them with the associated molecular emission.  The dust properties and mass are discussed in \S {4}.  The last section is a summary and comparison with the other clumps of knots. We suggest that the SW knots likely also arise from a similar set of high mass loss episodes.  

   \begin{figure}
   \figurenum{1}
   \epsscale{1.0}
   \plottwo{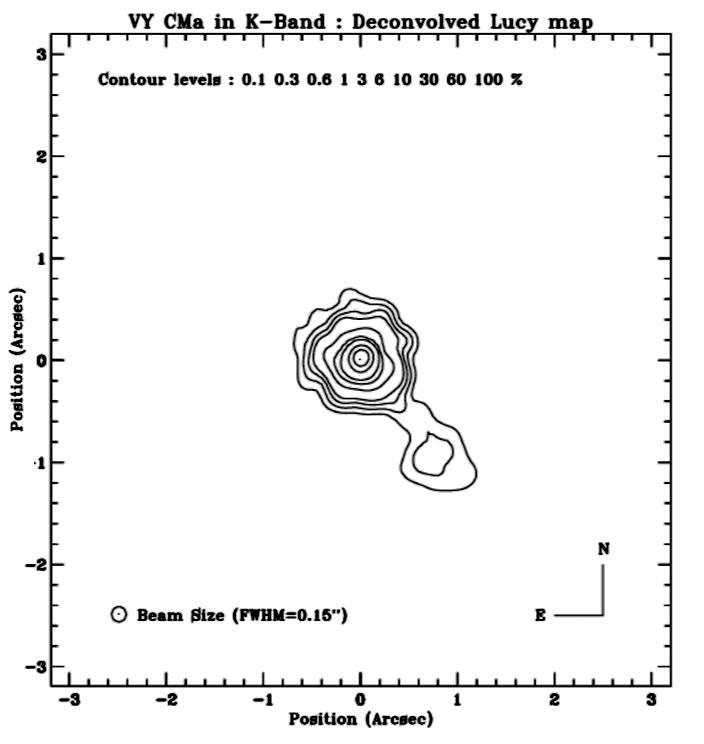}{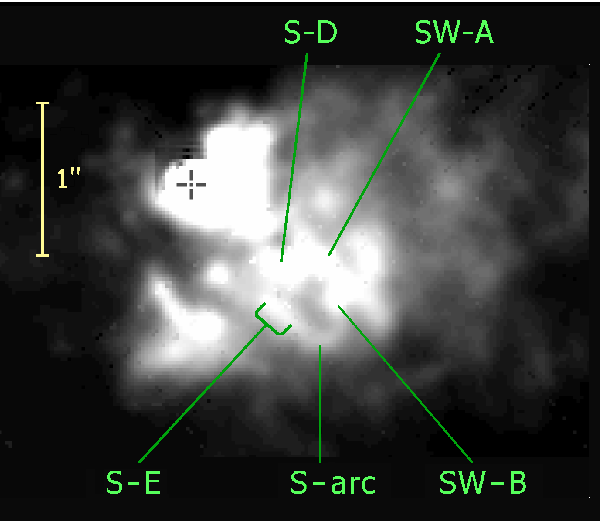}
	   \caption{Left: The  2.2$\mu$m K-band image of VY CMa from \citet{Cruz} obtained in 1994. Right: The HST F656N image observed in 1999 with the separate knots and arcs identified \citep{RMH2021}.  The + marks the position of the star.}
   \end{figure}

   \begin{figure}
   \figurenum{2}
   \epsscale{1.5}
   \plottwo{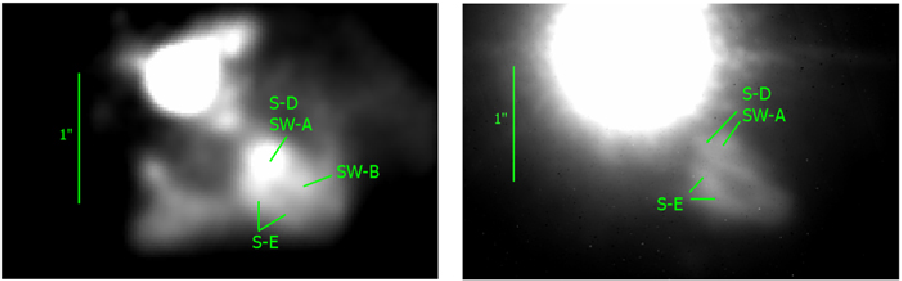}{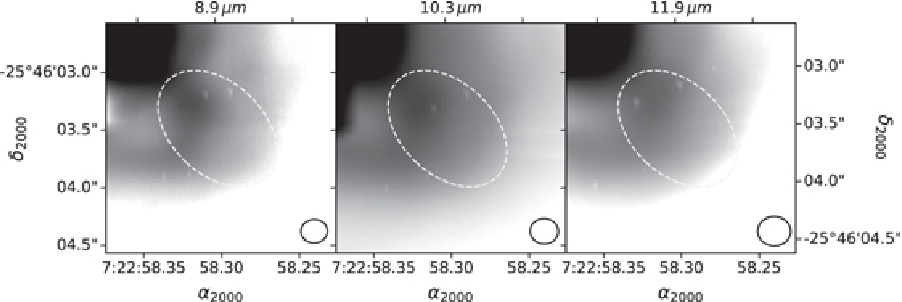}
   \caption{Top:The HST F1042 image (left) from 1999 with the visual knots identified, see Figure 1, and the K$_{s}$ image (right) from \citet{Shenoy13} with the knots marked. Bottom: The  9 -- 12 $\mu$m LBTI/NOMIC images from \citet{Gordon}.}  
   \end{figure}  

\section{The ALMA Observations }  

Our high resolution observations of VY CMa with ALMA, project 2021.1.01054.S, are described in \citet{Singh} and in the Appendix in \citet{RMH2024}.  The Band 6 data covering 216 to 270 GHz with a spatial resolution of 0\farcs2 and velocity resolution of 1.25 km s$^{-1}$ includes numerous molecular transitions. Ten are listed in Table B1 in \citet{Singh}.  The SW knots are identified in the molecular transitions listed here in Table 1.   Earlier observations have identified molecules in the SW region including TiO$_{2}$ \citep{debeck} and NaCl \citep{Decin}.


\input{Table1.tex}


     \vspace{2mm}  

In the following discussion we use the $^{12}$CO observations which have more complete coverage of the SW knots. 
Figure 3 shows the knots identified on channels from the  $^{12}$CO image cube corresponding with the knots on the HST images (Figure 1). The knots have different line of sight velocities. Hence they are not all visible in the same channel. The knots or condensations in the $^{12}$CO image cube appear more diffuse and appear more extended than in the HST images. The $^{12}$CO images have four times lower resolution, thus for comparison, we have applied an un-sharp masking code to the $^{12}$CO image cube and show three summed channels corresponding to the velocities of the respective knots marked in each image. When comparing directly with the HST image in Figure 1, note that nearly 23 years have passed between the HST observations and the ALMA image cube. The SW knots have significant and measurable 
transverse motion discussed in the next section.   

\vspace{2mm} 

The SW knots are spatially and kinematically different from the ALMA clumps described in \citet{RMH2024}. The ALMA clumps(A,B,C,D)  are more extended along the line of sight with broad velocity profiles that are more like extended ouflows than knots, especially clump D. The SW knots are more discrete with narrower profiles (see Appendix). Due to foreground extinction, the ALMA clumps are only visible at the mm wavelengths. There are no optical or infrared images for comparison. The ALMA clumps can be seen in  the upper left corner of Fig.3 very close to the star.   


    \begin{figure}
    \figurenum{3}
    \epsscale{1.0}
    \plotone{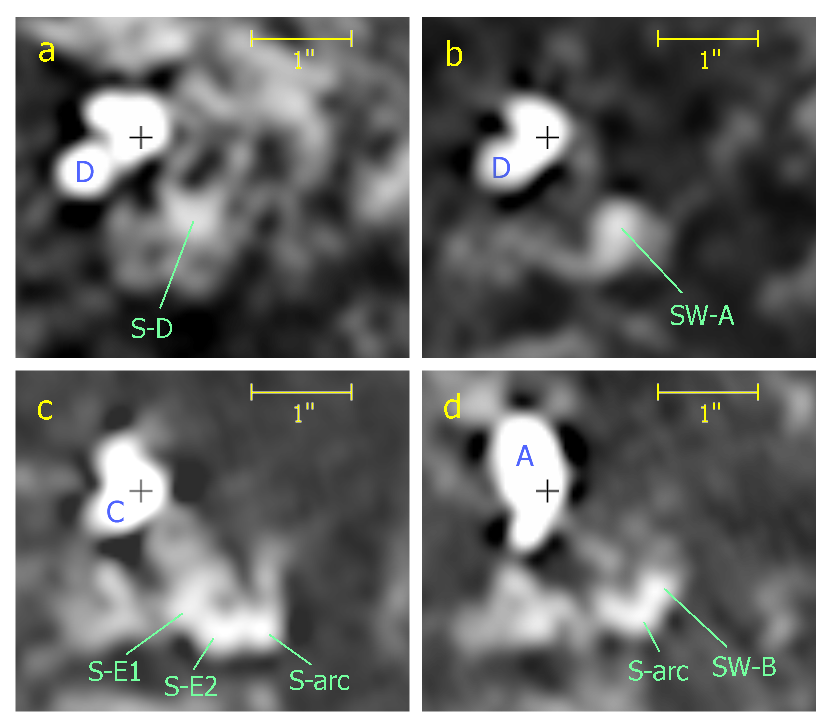}
	    \caption{Images of the SW knots from the $^{12}$CO image cube with the cross-identifications with the knots in the HST image, Figure 1. a. Knot S-D at V$_{LSR}$ 38 km s$^{-1}$.b. SW-A at V$_{LSR}$ 25 km s$^{-1}$. c. Knots S-E1, S-E2 and the S-arc in the arc that marks the east edge  of the SW knots, at V$_{LSR}$ 20 km s$^{-1}$. d.  SW-B at V$_{LSR}$ 10  km s$^{-1}$ also showing S-arc at this velocity. ALMA clump D is visible close to the star in images a and b, clump C in image c, and clump A in c and d to the north of the star.  The  + marks the position of the star in all the frames.}  
    \end{figure}


\section{The SW Knots -- Morphology and Motions }

The morphology of the SW knots, whether from a single high mass loss event or separate outflows over an extended period, depends on several parameters; their motions  relative to the star, the directions they are moving, orientation,  and the time since their ejection or age. 
\vspace{2mm}

In accordance with most papers that report
ALMA data, the Doppler line of sight velocities are in the LSR system. 

\vspace{2mm}

\subsection{Proper Motions and Ages}  

The proper motions with their errors are listed in Table 2  for each knot or condensation identified in Figure 1. Their corresponding direction of motion, ${\phi}$, measured N through E, and distance from the star in 1999 are also included. Their ages are estimated directly from their proper motions in arcsec per year and the distance from the star in arcsec \footnote{In our 2021 paper \citet{RMH2021}, we used the proper motions plus the line of sight velocity to derive the total velocity relative to the star, orientation, and distance from the star,  to estimate the age of each knot.  As discussed later in this section, the adopted velocity for VY CMa was probably affected  by passage through an optically thick cloud.} The proper motions were measured as described in \citet{RMH2007} from HST images in 1999 and 2005.  Since the reference frame for the motions is 1999, the ages are relative to that year.  


\input{Table2.tex}


\vspace{2mm}

The SW knots were all ejected from the star more than 200 years ago prior to the earliest record in VY CMa's historic light curve circa 1800.  See the light curves in  \citet{RMH2021}.  
Within the uncertainties,   except for SW-B, they are consistent with ages between  about 200 and 250 years.  As a group, the SW knots may be the result of an especially active period, similar to the outflows that coincide with VY CMa's 30 year active period in the early 20th century \citep{RMH2021,RMH2024}. Each structure is probably a separate event.

\subsection{The Line of Sight Velocities and Motion Relative to the Star} 

\vspace{2mm} 

The Heliocentric Doppler or line of sight velocities of the knots measured from the two  K I emission lines in the HST/STIS spectra \citep{RMH2021} are listed in Table 3. The LSR velocities of the $^{12}$CO emission at the corresponding knots or condensations are from the peak in their respective line profiles shown in the Appendix. The quoted uncertianties are from the FWHM of th profiles.  To compare with the  $^{12}$CO velocities, we correct the Heliocentric velocities for the motion of the Sun relative to the LSR \citep{Binney}. At the position of VY CMa this is -17.8 km s$^{-1}$ \footnote{The standard convention for ALMA observations uses an older value for the Solar motion of 20 km s$^{-1}$. Consequently, the velocity discrepancy could be uncertain by $\approx$ 2 km s$^{-1}$.}     


\input{Table3.tex}


       \vspace{2mm}

The Doppler velocities from the two independent sources, the K I emission and the $^{12}$CO molecular emission for knots S-D and SW-A, agree within the uncertainties, although the K I emission is redshifted relative to the S-E  feature and much moreso for the SW-B knot. The origin for this significant difference isn't understood. It may be due to distance from the star, the line of sight, differences in the excitation, or other factors. The similar line of sight velocities for the extended arc, 
S-E1, S-E2, and the S-arc indicates that they are part of a  physically  contiguous arc.  

\vspace{2mm} 

Table 4 includes the transverse velocity in the plane of the sky from the proper motions, and the line of sight velocity relative to the star (Vel$_{rel*}$) using  the $^{12}$CO Doppler velocities. To derive these values, we adopt the distance of 1.15 kpc and an LSR velocity of 22$\pm$1 km s$^{-1}$ for VY CMa, as discussed in \citet{RMH2024}.  In our 2021 paper \citep{RMH2021} we used a velocity for the star based on the velocity of the K I emission lines in an HST/STIS spectrum directly on the star. The $^{12}$CO observations, however, reveal extended emission in front of the star, slightly offset from the star's position, and with a  small blue shift from the star's velocity. See Figures 2 and 10 in \citet{RMH2024}. VY CMa faded 2.5 mags, a factor of 10, in the visual in the late 19th century,  as a result of one of its numerous outflow events\footnote{See the light curve Figure 9 in \citet{RMH2021}. The fading occurred between 1890 and 1900.}.  It has not brightened since. So it still obscured by the ejecta from that event. We suggest  that this $^{12}$CO material is the remnant from that outflow. We do not know how the optical emission  and absorption lines from VY CMa would be affected by passage through such dense and dusty material. For that reason we adopt the velocity of VY CMa from the radio observtions. Table 4 also gives the total velocity, the orientation or projection angle ($\theta$) of the knots with respect to the  plane of the sky, and their distances from VY CMa, measured in the 1999 HST images, corrected for the projection.      

         \input{Table4.tex}

\vspace{2mm} 

Three of the SW knots have projection angles near the plane of the sky.  Although the errors are large, especially for the S-E positions, this orientation  is consistent with the high polarization found for the SW knots \citep{Shenoy15} that requires an optimum scattering angle of 90 degrees, near the plane of the sky.

\subsection{Comparison with the $^{12}$CO Images and a Possible  Position Offset}

Its almost 23 yrs between the HST 1999 images and when the $^{12}$CO ALMA observations were obtained. Figure 4 shows the $^{12}$CO contours at two different velocities corresponding to the Doppler velocities of knots S-D and SW-A on the left and the S-E feature and S-arc on the right.  The offset in $\approx$ 23 yrs is very obvious. With their measured proper motions and corresponding transverse velocities, it is not surprising that there is a significant offset in the respective positions of the knots relative to star. To compare the images of the knots at the two  epochs, we show the measured distances of the knots from the star in Table 5,  and compare the observed position in 22.8 yrs in the $^{12}$CO image with the knot's expected position based on its proper motion.  They do not coincide. The differences are small; equivalent to 2 to 3 pixels or on the order of one resolution element in the ALMA image,  but they are systematic.   

    \begin{figure}
    \figurenum{4}
    \epsscale{0.8}
    \plotone{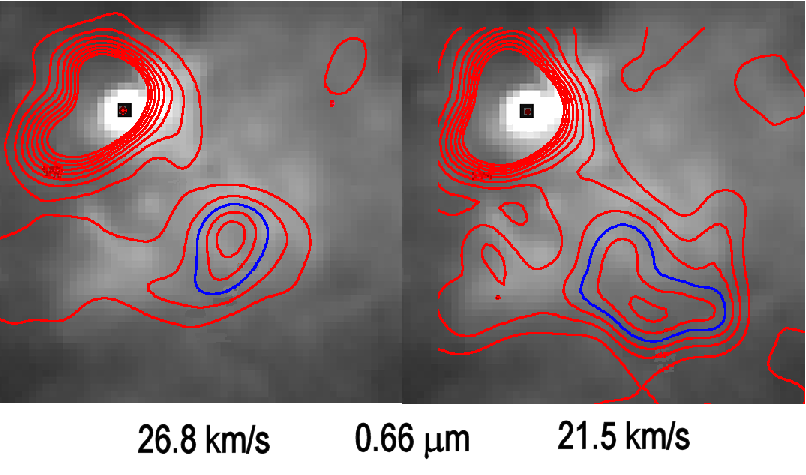}  
	    \caption{The $^{12}$CO contours in red on the F656N 1999 image from HST at two LSR velocities showing the offset in 22.8 yrs   for knots S-D and SW-A on the left and the S-E and S-arc on the right. The contours are linear with the 26.8 km s$^{-1}$ blue contour corresponding to 0.09 Jy/Beam and a spacing of 0.035 Jy/Beam; and the 21.5 km s$^{-1}$ blue contour corresponding to 0.1 Jy/Beam with a spacing of 0.02 Jy/Beam. The star's LSR velocity is 22 km  s$^{-1}$ . The small red spot in the black box marks the position of the star.}
    \end{figure}

\input{Table5.tex}

\vspace{2mm}

As a test we compared the expected position of the W2 knot \citep{RMH2019} with its observed position in the corresponding $^{12}$CO images. 
Its  projected distance from the star using its proper motion  agrees  with its distance from the star in $^{12}$CO\footnote{Knot W2 is 0$\farcs$484 from VY CMa in the 1999  HST images. With its proper motion 0.00377 $\pm$ 0.00081 arcsec/yr, its expected distance from the star in 22.8yrs is 0$\farcs$57 $\pm$ 0.02. Its $^{12}$CO distance is 0$\farcs$58 $\pm$ 0.02.} 

\vspace{2mm} 

We consider this systemetic difference between the observed and expected postions to be real. One possible explanation is acceleration of the ejecta in this direction relative to the star. For example, the offset for the three knots closest to the star, S-D, SW-A, and SW-B, would require an acceleration from $\approx$ 20 km s$^{-1}$ to 29 km s$^{-1}$ in 22.8 yrs or 0.4 km s$^{-1}$ yr$^{-1}$. This may seem small, but at distances more then 1000 AU, accleration by gas pressure, given the slow wind from VY CMa, would not be sufficient. Acceleration due to radiation pressure on dust and dust annealing has been suggested  (e.g. \citet{Chap}, \citet{RichardsY}) but this effect is unlikely to be sufficient at these distances. The difference is more likely due to the difference in the excitation mechanisms. The visual knots are due to scattered light while the CO emission is attributed to collisional and infrared excitation. Radiation from  the star will illuminate the outer side of the knot facing the star and the visual knot seen by reflection will appear to be somewhat closer. This could explain the rather small but systematic offset.

\vspace{2mm}

\section{The Dust Properties and Mass Estimates} 

In addition to the HST observations, the SW knots have been observed at near-infrared wavelengths from 2 to $5~\micron$ by \citet{Shenoy13}, in the $10~\micron$ band by \citet{Gordon}, and in the mm continuum (this paper). The emission at 2.2 and $3.6~\micron$ was pure scattered 
 light, and at $5~\micron$ the emission was mostly scattered light, with a 
modest contribution from thermal emission \citep{Shenoy13,Shenoy15}. Imaging 
polarimetry at 2.2 and $3.6~\micron$ \citep{Shenoy15} indicated a high fractional polarization and a position angle indicative of centro-symmetric scattering at these wavelengths. The emission in the $10~\micron$ band was purely thermal, as is the case at mm wavelengths. Gordon and Shenoy found that the SW knots 
were optically thick to scattering at near infrared wavelengths and optically thick in emission in the mid-infrared. They estimated a lower limit for the dust 
mass  
of M$_{dust}$ $>$ $5 \times 10^{-5}$M$_{\odot}$, or M$_{tot}$ $>$ $10^{-2}$M$_{\odot}$ with a gas-to-dust ratio of 200.

\vspace{2mm} 

By observing at wavelengths longer than $1~\micron$, we can penetrate deeper into 
the mass-loss wind of VY CMa, and see features that are hidden by dust extinction at shorter wavelengths.  If we can associate morphological features in infrared images of the SW Cump with CO emission,
we can better understand the 3D morphology deep into the circumstellar wind. The
 SW clump as seen in the infrared is not a single source, but has a more complex
 morphology with separate knots. In this section we will refer to the labeling 
in Figures 1 and 2. In Figure 5 we show the contours of CO emission at three 
different velocities overlaying the $3.6~\micron$ image. The CO emission at 26.8 km-s$^{-1}$ corresponds to the brightest $3.6~\micron$ emission which has the smallest 
projected distance from the star near knots SW-A and S-D. The CO emission at
18.1 km-s$^{-1}$ corresponds to the  emission with the greater projected 
distance from the star;
the S-E knots and the S-arc. The CO emission at 22.5 km-s$^{-1}$ straddles these
 two. 
Imaging polarimetry in the  R band with the HST \citep{jones2007} and in the near-infrard \citep{Shenoy15} are consistent with knot SW-A and the S-E knots lying close to the plane of the sky (Table 4).

\vspace{2mm} 

\begin{figure}
\figurenum{5}
\epsscale{1.0}
\plotone{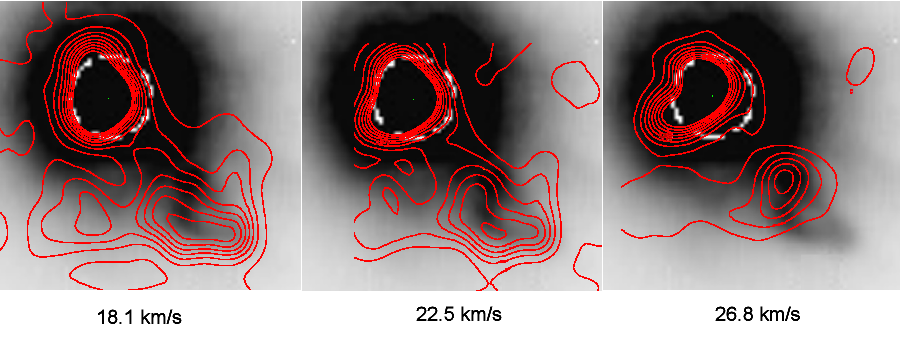}
\caption{CO emission contours at three different velocites are plotted 
	overlaying a grey scale image of the $3.6~\micron$ emission. The CO emission was averaged
 over a 4 km s$^{-1}$ bandpass for each plotted velocity. The contours are 
	linear, but arbitrarily chosen to emphasise the comparison between CO molecular emission and the scattered light seen at NIR wavelengths. The dashed white circle is the 
saturation level in the $3.6~\micron$. Each frame is $3 \arcsec$ wide, East to West.}
\end{figure}

A comparison of the near-infrared scattered light and the ALMA continuum 
emission is shown in Fig. 6. The mm continuum emission, associated with the S-E feature
 and S-arc identified in the HST and near-infrared images, is displaced from the
 $3.6~\micron$ emission to the southwest (S-E and S-arc), peaking about 
$2\arcsec$ from the star, or about 0.1 arcsec from the southwest tip of the $3.6~\micron
$ emission. Note that the primary peak in the $3.6~\micron$ emission (knots SW-A
 and S-D) 
is about $1.2\arcsec$ from the star, which is also the location of the peak in the mid-infrared thermal emission. There is only a modest increase in the mm continuum emission at this location. Some of the displacement can be attributed to the  motion of the features between 2011, when the near-infrared images were 
observed, and 2022, when the mm observations were made. The proper motions measured in the HST images  would account for about $0.05\arcsec$, roughly half the observed 
displacement. The lack of close correspondence of the mm dust emission with the 
mid-infrared thermal and near-infrared scattered emission is not clear, but a 
probable explanation can be made.

\vspace{2mm} 

\begin{figure}
\figurenum{6}
\epsscale{1.0}
\plotone{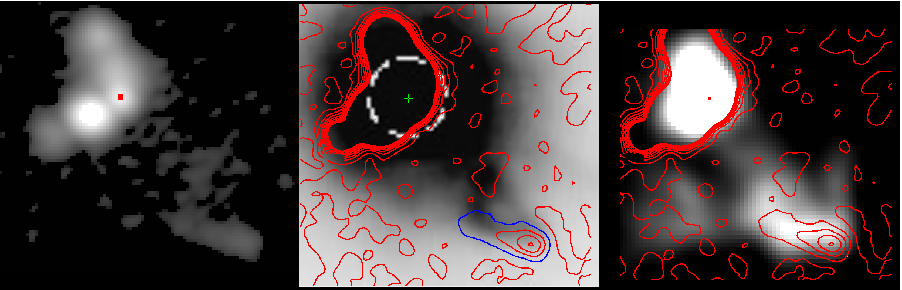}
\caption{Left Panel: Grey scale of the 249 GHz continuum emission.The large feature in the upper left includes the star and the ALMA clumps. Middle Panel:
 The 249 GHz continuum emission contours in red plotted overlaying a grey scale image of the 3.6$\mu$m emission. The contours are linear, but arbitrarily chosen to emphasize the comparison between mm wave dust emission and the NIR scattered emission.  Right Panel: The 249 GHz continuum emission contours plotted overlaying a grey scale image of the CO line emission integrated over $16-23$ km-s$^{-1}$. Each frame 
	is 3.1 arcsec wide, east to west. The small red dot or green cross marks the position of the star.}  
\end{figure}

The peak in the scattered and thermal infrared emission is likely closer to the 
star, where the dust is hotter and a more direct line-of-sight to the star is 
possible. \cite{Gordon} derived a dust temperature of T $\approx$ 200K at this 
location, requiring the 
9 to $12~\micron$ emission to come from the steep, Wien side of
the blackbody emission curve. The 249 GHz continuum brightness temperature at 
the NIR and MIR peak is $\approx 0.1$ K, indicating optically thin emission at 
mm wavelengths. Cooler dust further away from the star will show significantly lower emission at MIR wavelengths, but only moderately lower emission at mm 
wavelengths in the Rayleigh-Jeans regime (B $\propto$ T). The peak in the mm 
continuum emission just to the southwest of the NIR and MIR emission is indicative of the presence of significant dust mass that is likely blocked from a clear view of the star, reducing the scattered light. At a dust temperature of T $\lessapprox$ 150K, 
this location would also produce significantly less MIR emission as well. The 
solid angle subtended by the mm continuum emission is smaller than the NIR 
and MIR emission, so a greater column depth of dust at this location is possible. 
This could explain why the mm continuum is only weakly contributing
at the main peak in the infrared, where the column depth is lower, but spread 
out over a larger solid angle. This explanation suggests that there is an additional, significant dust mass associated with the SW knots a bit further to the southwest that is mostly not seen at near and mid-infrared wavelengths.

\vspace{2mm} 

In the mm-wave continuum, Clump C, Fig 6, upper left, \citep{OGorman, RMH2024}is very bright, but the SW knots,  which stand out in the infrared and optical red images, are relatively faint.  There are several reasons for this. First, Clump C is closer to the star and intercepts significantly more flux than the SW knots. Second, Clump C is embedded in the very dusty mass-loss wind from the star, and heavily obscured in the HST images and to a lesser extent in the NIR images \citep{Smith, Shenoy13}. Third,  the inner dust shell of VY CMa is so bright at NIR and Mid-IR wavelengths, that the exposures used to bring up the more distant regions in the wind severely overexpose the region near the star, in particular Clump C, even at $10~\micron$ \citep{Shenoy13, Shenoy15, Gordon}. It is possible that a series of much shorter exposures in the infrared using the adaptive optics systems on the LBT or the MMT could reveal the presence of Clump C. Note that while the details of the dust model, both composition and size distribution, can have an influence on differences between Clump C and the SW Clump, these are minor in comparison.

\vspace{2mm}

The flux density of the emission feature in the 249-GHz continuum associated 
with the S-arc, is 1.146 mJy in 0.35 arcsec$^{2}$, i.e. the blue contour in the middle panel of Figure 6. Following the method in \citet{OGorman} and adopted in \citet{RMH2024}, we can estimate the dust mass corresponding to this region. At a 
 distance of $1.93\arcsec$ from the star and a temperature of 150K, we get
 M$_d \approx 5.6\times 10^{-5}$ M$_\odot$ and total mass $1.1\times 10^{-2}$ M$ _{\odot}$. The dust mass estimate of M$_d$ $>$ $5\times 10^{-5}$ M$\odot$, or total mass M$_{tot}$ $>$ $10^{-2}$ M$_{\odot}$, from \citet{Shenoy13} and \citet{Gordon} 
applies primarily to knots SW-A and S-D at 1 to 1.2 $\arcsec$ from star. Thus, the mm continuum mass estimate is largely in addition to the mass lower limit derived from the NIR and MIR emission, yielding a total mass for the SW knots of 
M$_{tot}$ $>$ 2.1 $\times$ 10$^{-2}$ M$_{\odot}$.

\section{Summary}

The SW Clump or SW knots stand out in the ejecta of VY CMa. The group of knots and arcs is the brightest source of dusty infrared emission in addition to the central star itself. Initially, in the near-infrared, they looked like a separate feature or single ejection from VY CMa (Figure 1).  But the HST and ALMA images reveal three separate knots plus  an extended arc that are spatially and kinematically distinct.  

\vspace{2mm} 

The SW knots were ejected more than 200 years ago, so there is no record in VY CMa's historic light curve for comparison. Knot SW-A  and the S-E arc are consistent with ejection over a 30 year time span, but probably in separate outflows. Their motions with respect to the star indicate that they are essentially in the plane of the sky. Given the strong 1 to 12$\mu$m  flux from these same features and their common orientation, we suggest that they may have a nearly clear line of sight to the star's radiation. They may have been ejected from a common active region over a relatively short, 30 year, period of time, perhaps similar to VY CMa's active periods in the early 20th and mid-19th centuries.  A similar clear line of sight was suggested by \citet{RMH2019} to explain the very  strong K I emission lines observed in the knots just  west of VY CMa.

\vspace{2mm} 

The total mass for the SW knots, estimated from the infrared flux and from the continuum emission,  
is $>$ $2.1 \times 10^{-2}$ M$_{\odot}$. This is about half that measured for the obscured ALMA clumps just east of the star \citep{RMH2024}. We note that the mass estimates for the separate  knots in VY CMa's ejecta are typically 1 to 2 $\times$ 10$^{-2}$ M$_{\odot}$. This is obviously related to the energies involved and a clue to the mass loss mechanism.

\appendix
\section{Line Profiles for the knots}

Figure 7 shows the line profiles for the three distinct  knots or condensations, the two positions S-E1 and S-E2 and the S-arc.  
Measurements for each profile  were made at the  positions listed in Table A1. The scale of the $^{12}$CO frames is 0\farcs04/pixel where a pixel here is one unit in the x,y coordinate frame in each channel. 
In each relevant frequency channel, we simply measured the  
individual pixel value (Jy/beam) at each stated location.  
In this connection it is important to note that the spatial 
pixel width is entirely arbitrary in an inteferometric map,  
provided that it is much smaller than the resolution.  
Unlike most UV-to-IR data,  statistical noise does not manifest 
itself in pixel-to-pixel fluctuations.  
Instead each pixel represents the entire spatial beam 
centered at that location, and noise is spatially distributed.  
Hence the S/N would not be improved by taking a sum of multiple 
pixels centered at that location.  Moreover, in principle a larger 
sampling radius would slightly increase the amount of contamination 
by unrelated nearby spatial features.  In these data, a pixel is 0.04 
arcsec, compared    
to the 0.25 arcsec FWHM of the Gaussian beam produced by 
the inteferometry reduction procedure.

\input{TableA1}

 A related consideration involves spatial features that are not
fully resolved from each other.  If they are strongly correlated,
any similarity of their spectra naturally means little.  However,
an observed difference between them is significant even though
it may be difficult to quantify. 

\vspace{2mm} 

Since each location has several velocity components along its line of 
sight, we identified the separate knots based on the local spatial 
structures seen in the relevant ALMA frequency channels, see Figure 3.  
 The prominent emission feature near the velocity of the star in the S-D profile, for example, is due to extended emission, not the S-D knot.  At these velocities it may be an extension of the S-E arc in Figure 3c, and is likewise near the plane of the sky. S-D and SW-A have peak fluxes at very different frequencies although their projected separaton at 0\farcs25 is marginal. The same emission feature  can be seen in the profile for SW-A.   Knots S-E1 and S-E2 are on a common arc-like feature, but separated by more  than 0\farcs25 and are therefore separate measurements.  

\vspace{2mm} 

 All of the profiles, at different positions and distances from the star, show prominent emission peaks at $\pm$ 30 --35 km s$^{-1}$ relative to the star's LSR velocity at 22 km s$^{-1}$ marked by a vertical line. This is emission from older, outer ejecta at 3000 to 8000 AU, more than 800 yr old,
 that was described in Section 6 in \citet{RMH2024}. Note the steep decline or edge at the extreme ends of each profile.

    \begin{figure}
    \figurenum{7}
\epsscale{1.0}
\plotone{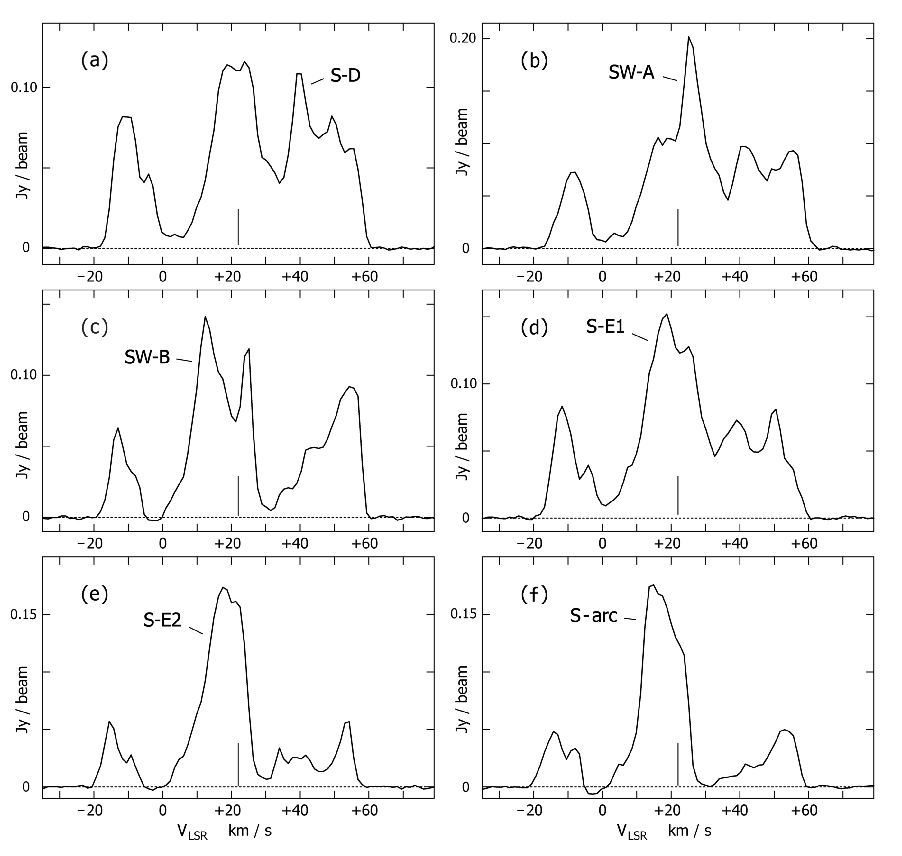} 
	    \caption{The continuum-subtracted $^{12}$CO line profiles. The short vertical line marks the velocity of the star at 22 km s$^{-1}$.}
    \end{figure}


\begin{acknowledgments}

This paper uses of ALMA data:ADS/JAO.ALMA$\#$2021.1.01054.S. 
ALMA is a partnership of ESO (representing its member states), NSF (USA), and NINS (Japan), together with NRC
(Canada) and NSC and ASIAA (Taiwan) and KASI (Republic of Korea),
in cooperation with the Republic of Chile. The Joint ALMA Observatory is
operated by ESO, AUI/NRAO, and NAOJ. The National Radio Astronomy
Observatory is a facility of the National Science Foundation operated under
cooperative agreement by Associated Universities, Inc. 

HST images of VY CMa are used in this paper. The data can be found at
\dataset[DOI: 10.17909/brdn-2720]{https://doi.org/DOI: 10.17909/brdn-2720[doi.org]}. Dataset Title: WFPC2 images of VY CMa DOI: 10.17909/brdn-2720.

This research is supported by NSF Grants AST-1907910 and AST-2307305. 

\end{acknowledgments}

\end{document}

%% file: Table1.tex
\begin{deluxetable}{lll} 
\tablewidth{0 pt}
\tabletypesize{\footnotesize}
\tablenum{1} 
\tablecaption{Molecular Transitions Between 216--270 GHz Identified in the SW Knots}
\tablehead{
\colhead{Molecule} & 
	\colhead{Transition}  &
\colhead{SW Knots}   
}
\startdata
	$^{12}$CO  &   J $=$ 2 $\rightarrow$ 1  &  All \\
	$^{13}$CO  &   J $=$ 2 $\rightarrow$ 1  &  S-D/SW-A, S-E1/E2 S-arc \\
	NaCl       &   J $=$ 18 $\rightarrow$ 17 & S-E1/E2, S-arc \\
        SO$_{2}$   &   J$_{ks,kc}$ $=$ 13$_{1,13}$ $\rightarrow$ 12$_{0,12}$ & S-E1/E2, S-arc \\ 
	PN\tablenotemark{a}  &   J $=$ 5 $\rightarrow$  4  & S-D, SW-A,SW-B, S-E1/E2 \\
        SiO\tablenotemark{a} &   J $=$ 5 $\rightarrow$ 4   & SW-B, S-E1/E2, S-arc             
\enddata
	\tablenotetext{a}{See \citet{ravi}} 
\end{deluxetable} 

%% file: Table2.tex
\begin{deluxetable}{lllll} 
\tablewidth{0 pt}
\tabletypesize{\footnotesize}
\tablenum{2} 
\tablecaption{Proper Motions and Ages for the SW Knots}
\tablehead{
\colhead{Knot} & 
	\colhead{Proper Motion}  &
	\colhead{Direct. $\phi$} &
	\colhead{Distance} &
	\colhead{Age}\\
	&
	\colhead{arcsec/yr}  &
	\colhead{Degrees}  &
	\colhead{Arcsec} &
	\colhead{years} 
}
\startdata
	S-D  &  0.00296 $\pm$ 0.00080 & -152 &   0.86  &    290 $\pm$ -60,+119   \\
	SW-A &  0.00424 $\pm$ 0.00052 & -101 &   0.96  &    226 $\pm$ -24,+32 \\
	SW-B &  0.00373 $\pm$ 0.00133 & -137 &   1.24  &    332 $\pm$ -87,+185  \\
	S-E1 &  0.00408 $\pm$ 0.00129 & -117 &   0.99  &    236\tablenotemark{a}  $\pm$ -52,+123  \\  	
	S-E2 &  0.00584 $\pm$ 0.00175 & -129 &   1.14  &    197 $\pm$ -47,+81     
\enddata
	\tablenotetext{a}{S-E1 and S-E2 are two positions along the diffuse arc that marks the eastern edge of the SW knots or the side of a loop including the S-arc in Figure 1.  The measurements for S-E1 and S-E2 thus represent the same structure with an avergae age of 216 yrs.} 
\end{deluxetable} 

%% file: Table3.tex
\begin{deluxetable}{lllll} 
\tablewidth{0 pt}
\tabletypesize{\footnotesize}
\tablenum{3} 
\tablecaption{Line of Sight Velocities for the SW Knots}
\tablehead{
\colhead{Knot} & 
	\colhead{ Vel$_{Hel}$(K I) } &
	\colhead{ Vel$_{LSR}$(K I) } &
	\colhead{ Vel$_{LSR}$($^{12}$CO) } &
	\colhead{ Diff.\tablenotemark{a} } \\ 
\colhead{}	&
	\colhead{ km s$^{-1}$ }  &
	\colhead{ km s$^{-1}$ }  &
	\colhead{ km s$^{-1}$ }  &
	\colhead{ km s$^{-1}$ }  
   }
\startdata
	S-D  &  65.1$\pm$0.6  & 47.3  & 40.3$\pm$4.5 &  +7    \\
	SW-A &  48.0$\pm$0.9  & 30.2  & 25.0$\pm$4   &  +5    \\
	SW-B &  63.5$\pm$0.7  & 45.7  & 12.4$\pm$5   & +33    \\  
	S-E1 &  53.0$\pm$0.2  & 35.2  & 18.7$\pm$9   & +17.5  \\  	
	S-E2 &  53.2$\pm$0.7  & 35.4  & 17.5$\pm$7   & +17.9  \\    
	S-arc & \nodata       & \nodata & 15.0$\pm$8   & \nodata  
\enddata
\tablenotetext{a}{ Difference,  Vel$_{LSR}$(K I) - Vel$_{LSR}$($^{12}$CO) }
\end{deluxetable} 

%% file: Table4.tex
\begin{deluxetable}{llllll} 
\tablewidth{0 pt}
\tabletypesize{\footnotesize}
\tablenum{4} 
\tablecaption{Velocities, Orientation, and Distances}
  \tablehead{
    \colhead{Knot}   & 
    \colhead{Trans.Vel.}   &
	\colhead{Vel$_{rel*}$}\tablenotemark{a}  &  
    \colhead{Total Vel.}    &
    \colhead{Orient $\theta$} &
    \colhead{Dist. 1999} 
         \\ 
    \colhead{}             &  
    \colhead{km s$^{-1}$}  &  
    \colhead{km s$^{-1}$ rel*}  &  
    \colhead{km s$^{-1}$}  &  
    \colhead{degrees}      & 
    \colhead{AU}   
  } 
\startdata
S-D & 16 $\pm$ 4.4  &  18 $\pm$ 4.5 & 24 $\pm$ 6  &  +48 $\pm$ 10.7 &  1476  \\
SW-A & 23 $\pm$ 2.8 &  3 $\pm$ 4 & 23 $\pm$ 5 & +7.4 $\pm$ 9.8  & 1112   \\ 
SW-B & 20 $\pm$ 7.2 &  -10 $\pm$ 5  &  22 $\pm$ 8 & -26.5 $\pm$ 14  &  1593  \\
S-E1  & 22 $\pm$ 7.0  & -3 $\pm$ 9  &  22 $\pm$ 11  &  -7.8 $\pm$ 23  &  1148 \\
S-E2  & 32 $\pm$ 9.5 & -4.5 $\pm$ 7 & 32 $\pm$ 12  &  -8.0 $\pm$ 12.3 &  1323 \\
S-arc & \nodata & -7 $\pm$ 8 &  \nodata & \nodata &  1586  
\enddata
	\tablenotetext{a}{The star's LSR velocity is 22 km s${-1}$}
\end{deluxetable} 

%% file: Table5.tex
\begin{deluxetable}{lcccc} 
\tablewidth{0 pt}
\tabletypesize{\footnotesize}
\tablenum{5} 
\tablecaption{Positional Offset 22.8 yrs}
\tablehead{
\colhead{Knot} & 
	\colhead{Dist. 1999.2} &
	\colhead{Dist. CO 2021.96 \tablenotemark{a}} &
	\colhead{Expected Dist. 22.8yrs} &
	\colhead{Diff.\tablenotemark{b}} \\ 
        \colhead{}	&
	\colhead{arcsec}  &
	\colhead{arcsec}  &
	\colhead{arcsec} &
	\colhead{arcsec}
}
\startdata
S-D  &  0.86 & 1.04  & 0.93 & 0.11 \\
	SW-A &  0.96 & 1.20  & 1.07 & 0.13   \\
	SW-B &  1.24 & 1.43  & 1.31   & 0.12  \\  
	S-E1 &  0.99 & 1.22 & 1.06 & 0.15  \\  	
S-E2 &  1.14 & 1.55 &  1.28  & 0.27 \\    
	S-arc & 1.38 & 1.73  & \nodata  & \nodata  
\enddata
       \tablenotetext{a}{The uncertainty in the CO positions is $\leq$ 0.04 arcsec or approximnately 1 pixel.}
	\tablenotetext{b}{Difference - The $^{12}$CO observed distance from the star minus the expected distance in 22.8 yrs from the measured proper motions. The $^{12}$CO positions are those used for the line profile measurements (Appendix A).}  
\end{deluxetable} 

%% file: TableA1.tex
\begin{deluxetable}{lllll}
\tablewidth{0 pt}
\tabletypesize{\footnotesize}
\tablenum{A1} 
	\tablecaption{ $^{12}$CO Measurement Positions and Distance from Star\tablenotemark{a}}
\tablehead{
\colhead{Knot} & 
	\colhead{Channel} &
	\colhead{x position} &
	\colhead{y position} &
	\colhead{Dist from star} 
}
\startdata
	S-D  &  66 & 465.25 & 433.75 & 1\farcs04  \\
	SW-A &  76 & 471.75 & 433.25 & 1\farcs20  \\
	SW-B &  89 & 480.5  & 434.0  & 1\farcs43 \\  
	S-E1 &  82 & 465.5  & 429.5  & 1\farcs22  \\  	
	S-E2 &  83 & 473.5 & 423.5  & 1\farcs55  \\     
	S-arc & 85   &  478.5 &  422.0 & 1\farcs73
\enddata
	\tablenotetext{a}{The star is at  x 453.51, y 457.23 (7:22:58.32, -25:46:03.01 \citet{RMH2024}); the scale is 0\farcs04/pixel.}
\end{deluxetable} 